\documentclass{article}
\usepackage{spconf,amsmath,graphicx}
\usepackage{hyperref}
\usepackage{url}
\usepackage{wrapfig}
\usepackage{caption}
\usepackage{float}
\usepackage{xcolor}
\usepackage{multirow}
\usepackage{graphicx}

\title{HumTrans: A Novel Open-Source Dataset \\ for Humming Melody Transcription and Beyond}
\name{Shansong Liu$^{1\dag}$, Xu Li$^{1}$, Dian Li$^{2\dag}$, Ying Shan$^{1}$
\thanks{${\dag}$ Corresponding authors.}
}
\address{
    $^1$ ARC Lab, Tencent PCG \\
    $^2$ Foundational Technology Center, Tencent PCG
}

\begin{document}
\ninept
\maketitle
\begin{abstract}
This paper introduces the \textit{HumTrans} dataset, which is publicly available and primarily designed for humming melody transcription. The dataset can also serve as a foundation for downstream tasks such as humming melody based music generation. It consists of 500 musical compositions of different genres and languages, with each composition divided into multiple segments. In total, the dataset comprises 1000 music segments. To collect this humming dataset, we employed 10 college students, all of whom are either music majors or proficient in playing at least one musical instrument. Each of them hummed every segment twice using the web recording interface provided by our designed website\footnotemark[1]. The humming recordings were sampled at a frequency of 44,100 Hz. During the humming session, the main interface provides a musical score for students to reference, with the melody audio playing simultaneously to aid in capturing both melody and rhythm. The dataset encompasses approximately 56.22 hours of audio, making it the \textbf{largest known humming dataset to date}. The dataset will be released on Hugging Face\footnotemark[2], and we will provide a GitHub repository containing baseline results and evaluation codes\footnotemark[3].
\end{abstract}

\footnotetext[1]{\href{https://www.humming-collect.online}{HumTrans Dataset Collection Website}}
\footnotetext[2]{\href{https://huggingface.co/datasets/dadinghh2/HumTrans}{HumTrans Dataset Hugging Face Repository}}
\footnotetext[3]{\href{https://github.com/shansongliu/HumTrans}{HumTrans Dataset GitHub Repository}}
%
\section{Introduction}
\label{sec:intro}

Music transcription refers to the process of converting instrumental music or vocal sounds such as singing, humming, or whistling into musical notations \cite{ryynanen04_sapa,poliner2007melody}. Conventionally, music transcription has been a manual task requiring trained individuals with a background in music education. Even for professionals, achieving error-free transcriptions is not trivial. It often requires repetitive listening to the music pieces, which is extremely time-consuming. An alternative approach is to develop automatic music transcription (AMT) algorithms \cite{klapuri2009automatic,benetos2013automatic,benetos2018automatic,wu2021omnizart,gardner2022mt,donahue2022melody,yong2023phoneme} which can bring significant benefits to music professionals and users seeking quick access to sheet music for specific music pieces. Moreover, such functionality can be integrated into consumer applications, enabling users to input an audio or their own voice to obtain the corresponding music score as output for further use.

A bunch of early AMT approaches relied on signal processing techniques such as autocorrelation \cite{brown1991musical,brown1993determination, monti2000monophonic} and spectral or cepstral features \cite{martin1996blackboard,goto2001predominant,clarisse2002auditory} to track pitch information. These two categories of methods tackle the problem of AMT from the perspectives of time domain and frequency domain, respectively. Autocorrelation based methods are relatively robust to noise but sensitive to formant structures, while spectral or cepstral based methods perform less effectively in noisy environments \cite{klapuri2004signal}. Recognizing the limitations of signal processing based AMT models, researchers moved to Hidden Markov Model (HMM) based methods \cite{ryynanen2005polyphonic,ryynanen2006transcription,poliner2006discriminative}. For example, the authors of \cite{poliner2006discriminative} employed a pitch-wise HMM to represent note activity and inactivity, where the HMM parameters were learned from a training set with ground-truth pitch information. A multi-pitch detection system proposed by \cite{ryynanen2005polyphonic} used a 3-state HMM to represent a note event. Nevertheless, HMM based AMT methods may struggle with capturing the rich diversity and variability of music signals since music encompasses various dynamics, timbres and expressive nuances which are not easily captured by the simplified probabilistic models used in HMMs.

In recent years, neural network based methods have been widely applied in the AMT field \cite{sigtia2016end,roman2018end,wu2020multi,gardner2022mt} due to their superior ability in capturing complex patterns and relationships in music data. Sigtia et al. \cite{sigtia2016end} presented a network model for polyphonic piano music transcription. The model combines an acoustic model, which estimates pitch probabilities in audio frames, with a music language model that captures pitch correlations over time. Rom{\'a}n et al. \cite{roman2018end} proposed an end-to-end framework for audio-to-score music transcription without separate stages for pitch detection and note tracking by utilizing a convolutional recurrent neural network. A notable research in the AMT field is the MT3 \cite{gardner2022mt} which employed an off-the-shelf T5 model \cite{raffel2020exploring} on combined datasets with different instrument types and outperformed the current state-of-the-art (SOTA) models on each individual dataset.

\begin{figure*}[htbp]
\centering
\includegraphics[width=17cm]{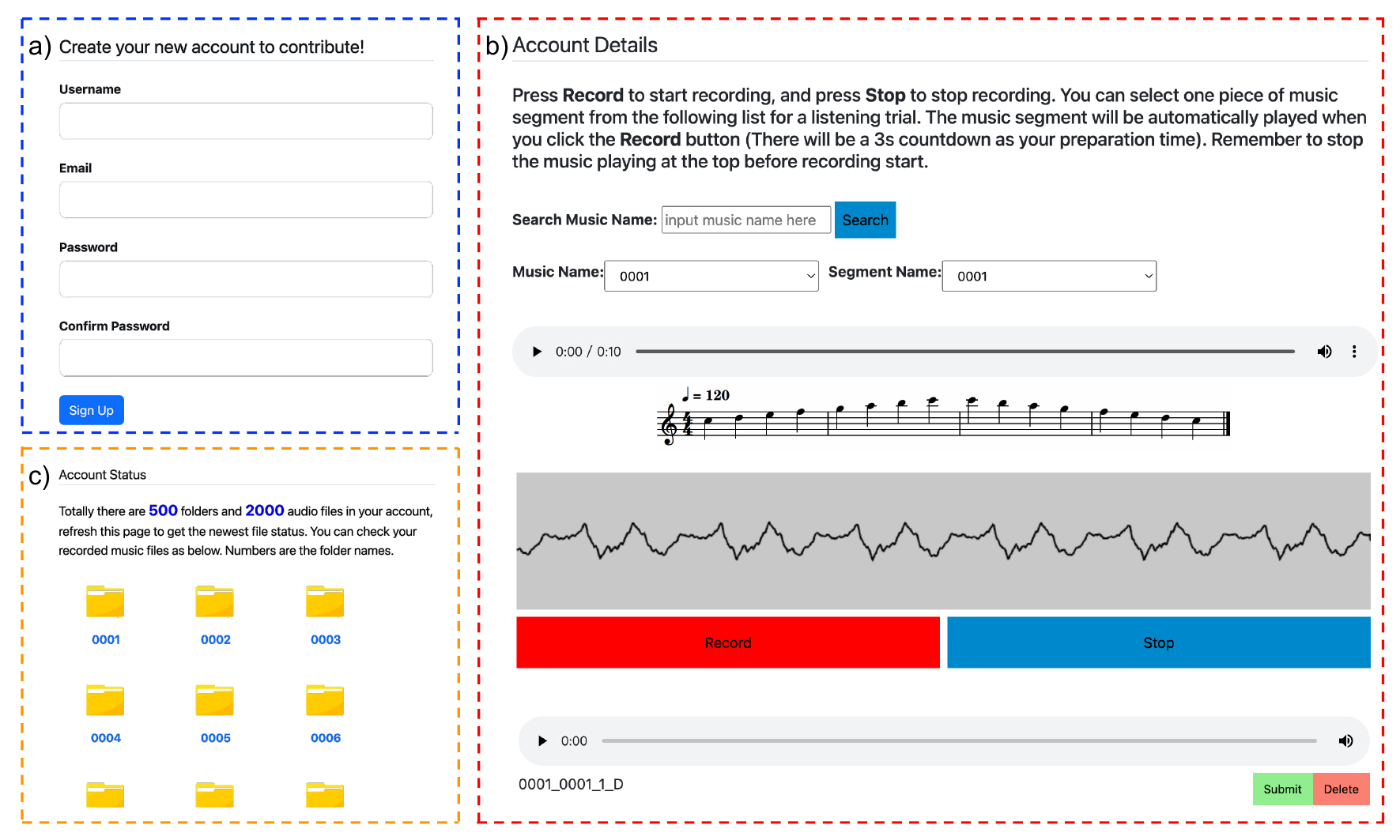}
\caption{HumTrans dataset collection web interface. a) user registration module; b) main interface for humming data recording; c) a data monitor module for users to access to their recording status.}
\label{fig:humtrans_web}
\end{figure*}

However, the majority of existing AMT researches has primarily focused on instrumental data, while there is relatively less research on AMT for vocals \cite{ryynanen2006transcription,rigaud2016singing,donahue2022melody,wang2022musicyolo}. Moreover, there is a notable absence of representative works similar to MT3 \cite{gardner2022mt} in this context. One possible reason is that vocals are comparatively more challenging to transcribe accurately, as they exhibit less stability, and pitch transitions are not as sharp as instruments. Acquiring and labelling vocal data for note transcription is also more difficult compared to instruments, as humans do not have the same direct control ability over producing specific pitches as instruments do, and even professional singers often rely on instruments to calibrate their pitch. In the AMT field of vocal transcription, there is a limited amount of available singing data \cite{wang2021preparation,wang22b_interspeech,zhang2022m4singer}, while humming data is extremely scarce. There is a few humming data released by the MIREX challenge, but the dataset is small ($<$10 hours) and the curation is insufficient (dataset link unavailable on the website). Therefore, there is a pressing need to collect a humming dataset for public use.

In this paper, we introduce the \textit{HumTrans} dataset. To the best of our knowledge, this is the largest available humming dataset to date consisting of solely hummed melodies. In this dataset, 10 college students were employed and individually hummed 1000 music segments, with each student either majoring in music or proficient in at least one musical instrument. All the humming data was recorded by our designed web interface with a sampling rate of 44,100 Hz. The total duration of the recordings is around 56.22 hours. The dataset will be released on Hugging Face with associated baseline results and evaluation codes stored in a GitHub repository.

The rest of the paper is organized as follows. Section \ref{sec:humtrans_collect} presents our designed web interface and the process of collecting the humming data. Section \ref{sec:humtrans_analysis} provides a statistical analysis of the humming dataset. Baseline results will be shown in Section \ref{sec:baseline}. Finally, Section \ref{sec:conclusion} concludes this paper.

\section{Creation of the HumTrans Dataset}
\label{sec:humtrans_collect}

\noindent
To collect the humming dataset, we developed a dedicated web interface (link provided in the footnote on the first page). The interface consists of three main modules: a user registration module, a main interface for humming recording, and a module to monitor the user's current recording status. The source code for building the web interface will also be open sourced\footnotemark[4], allowing researchers to modify it according to their own needs and requirements.

\footnotetext[4]{\href{https://github.com/shansongliu/HumTransDatasetWeb/}{HumTrans Dataset Web Interface Source Code}}

\subsection{HumTrans Dataset Collection Web Interface}
\label{subsec:humtrans_web_interface}

\subsubsection{User Registration Module}
\label{subsubsec:humtrans_user}

\noindent
Developing a user registration module facilitates the segregation of recordings from different users and also streamlines the processing of data from a specific user (as shown in Fig. \ref{fig:humtrans_web}a). When users need to record their humming data, they can easily start the recording process by logging into the website using their registered username and password.

\subsubsection{Main Recording Interface}
\label{subsubsec:humtrans_record}

The main recording interface is shown in Fig. \ref{fig:humtrans_web}b. The general process is as follows. Users first select a segment from the melody pool and then hum the melody based on the chosen one. After finishing the humming, the recorded segment will be displayed at the bottom of the page, where users can playback, delete, or upload the recorded segment.

The recording interface features two functions for users to choose the music segment. One is the ``Search Music Name'', which allows users to input a music ID to navigate to that musical composition. The other function is the drop-down menus. Users can first select a music ID in the left menu and then pick the segment of that music ID in the right menu. The audio of the chosen music segment is presented under the drop-down menus for auditioning. The associated sheet music is displayed as well to assist users in familiarizing themselves with the melody they are about to hum.

Above the ``Record'' button, there is a grey canvas that captures the external sound in real-time through the microphone, demonstrating the waveform of the incoming sound. Users can observe the waveform to assess if there is any noise in the current environment. If the noise is significant, they can choose to stop recording or change their recording environment. By clicking the ``Record'' button located at the bottom left of the canvas, the system will provide a 3-second beep sound to allow users to prepare themselves before humming. Once the beep sound ends, users need to hum along with the melody playing through their headphones. After completing the humming, users can click ``Stop'' to finish the recording and enter a file name. The recorded audio segment will be presented at the bottom of the page. Pressing the ``Submit'' button the audio segment will be submitted to our system.

\subsubsection{Data Monitor Module}
\label{subsubsec:humtrans_monitor}

\noindent
The purpose of the data monitoring module is to provide users with information on the current number of recorded segments, as depicted in Fig. \ref{fig:humtrans_web}c. The top of the page shows the number of folders and the total count of recorded segments. Each folder represents a musical composition. Users can enter the folders to view specific recorded humming segments, and if they find a recorded segment unsatisfactory during playback, they have the option to delete and record the segment again.

\subsection{Process of the HumTrans Dataset Collection}
\label{subsec:humtrans_process_data}

\subsubsection{Initialization of the Melody Pool}
\label{subsubsec:humtrans_melody_pool}

We initially gathered a collection of publicly available sheet music (approximately 5000 pieces) from the internet. We listened to the corresponding audio for each piece of sheet music and excluded those that were unsuitable for humming. In the end, we narrowed it down to 500 musical compositions. ``Unsuitable for humming'' refers to cases where the pitch is too high, exceeding the vocal range of males and females, or situations where there are significant intervals between adjacent notes, making it challenging for human vocals to handle, such as certain piano pieces.

For each collected sheet music, we used the notation creation function of an open-source software MuseScore 4\footnotemark[5] to make electronic sheet music. We then utilized the built-in conversion function of MuseScore 4 to convert the electronic sheet music into MIDI format for further use. Each sheet music was divided into 1 to 4 segments, resulting in 1000 MIDI files. A unique identifier was assigned to each MIDI file based on the music ID and segment ID, with the music ID ranging from 0001 to 0500 and the segment ID ranging from 0001 to 0004. The obtained segments only include the melodic parts of the sheet music, discarding the chord accompaniment. Additionally, we inserted a few rests at appropriate positions to facilitate users to take breaths while humming. Subsequently, we exported the MIDI files into MP3 format and uploaded them to our system as reference audios for humming.

\footnotetext[5]{\href{https://muse3core.org/en/4.1}{MuseScore 4}}

\subsubsection{Recording and Processing of the Data}
\label{subsubsec:humtrans_record_process}

\noindent
Prior to recording, the students were asked to inspect their devices to guarantee that the built-in microphone was functioning properly. We requested that students wear headphones while humming to ensure that the melody being played simultaneously would not be recorded. Additionally, if there was excessive ambient noise, the students were responsible for finding another relatively quiet environment. After recording, they were required to check the quality of each recorded segment. If there were instances of humming out of tune or the students found it difficult to keep up with the rhythm of the melody played during the recording session, it was necessary for the students to record the segment again.

We conducted a further sanity check to make sure that each recorded file name had the ``musicID\_segmenID'' prefix. In the suffix (``\_1\_D'') shown in the file name of Fig. \ref{fig:humtrans_web}b, ``1'' represents the first recording of segment ``0001\_0001'', while ``2'' can  indicate the second recording of the same segment. Letter ``D'' signifies that students recorded the segment by lowering one octave based on their vocal range. If it is ``DD'', it means they lowered their voice by two octaves, and so on. All recorded files were scored by DNSMOS \cite{reddy2021dnsmos} to evaluate recording quality. Those with a quality score below 2.8 were excluded. Finally, 14,614 recording files were obtained with a total duration of 56.22 hours.

\begin{figure}[htbp]
\centering
\includegraphics[width=8.7cm]{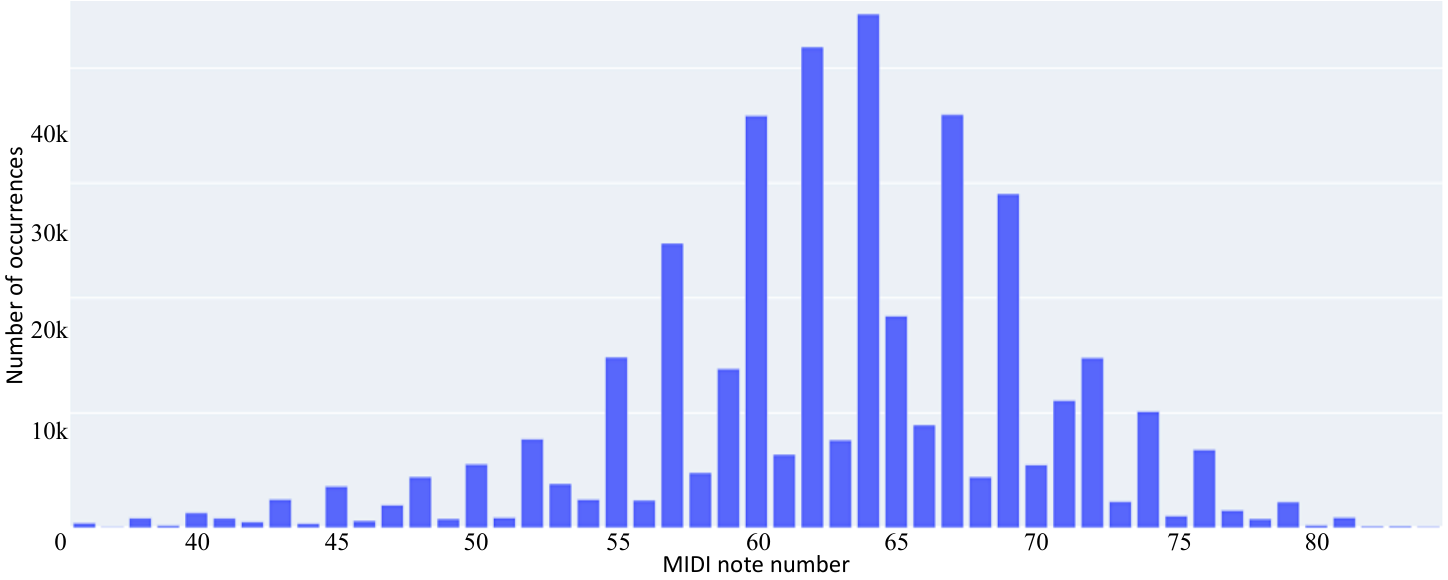}
\caption{The statistical distribution of pitch. Pitch is presented in the form of MIDI note numbers, where 60 is C4.}
\label{fig:humtrans_pitch_distribution}
\end{figure}

\section{Statistics of HumTrans Dataset}
\label{sec:humtrans_analysis}

\begin{figure}[htbp]
\centering
\includegraphics[width=8.7cm]{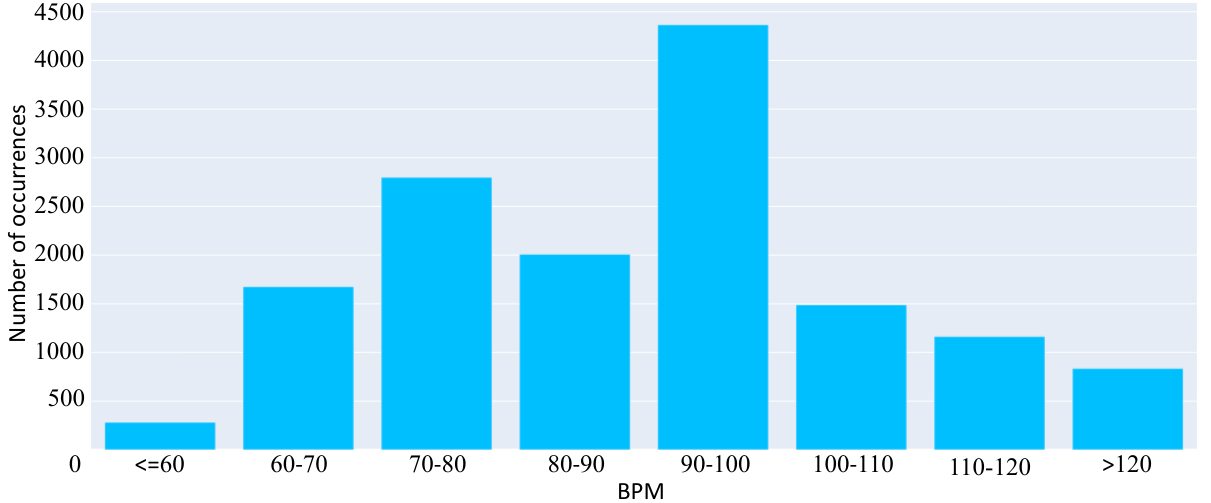}
\caption{The statistical distribution of BPM.}
\label{fig:humtrans_bpm_distribution}
\end{figure}

\noindent
To collect a gender-balanced humming dataset, we recruited 10 students, 5 males and 5 females, who are either majoring in music or proficient in at least one musical instrument. They have a range of both high and low voices, resulting in a wide vocal range when combined, spanning five octaves from low C (C2) to high C (C6). Therefore, the dataset encompasses a rich variety of pitches, as shown in Fig. \ref{fig:humtrans_pitch_distribution}. It can be observed that the pitch range is mainly distributed from 57 (A3) to 69 (A4), which is also the region where male and female vocal ranges overlap more. The selected melody pool contains a higher frequency of pitches from the white keys (C/D/E/F/G/A/B) of the piano, while the pitches from the black keys (e.g., G\#) are less frequent. Our HumTrans dataset covers a wide range of BPM as well, from 52 to 156, as shown in Fig. \ref{fig:humtrans_bpm_distribution}, enabling the humming transcription system trained on our dataset to handle various BPM conditions.

\begin{figure}[htbp]
\centering
\includegraphics[width=8.7cm]{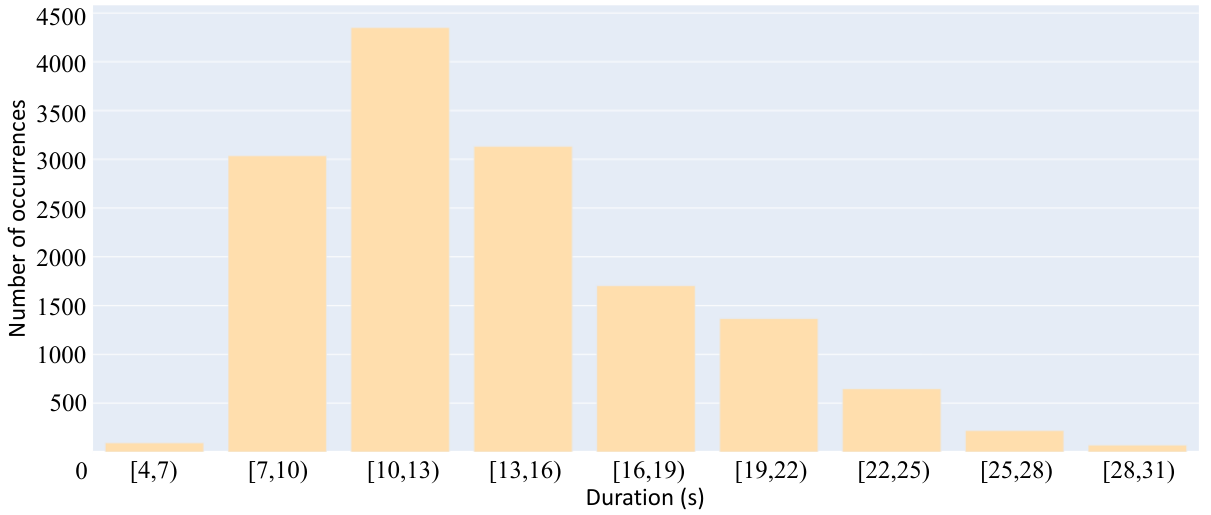}
\caption{The statistical distribution of each music segment.}
\label{fig:humtrans_dur_distribution}
\end{figure}

We asked the students to utilize the syllables ``Da-Da-Da'' during the recording to represent humming. Additionally, they were instructed to synchronize their humming with the rhythm of the played melody. This approach ensured that our data is self-labeled, eliminating the need for additional manual annotation. As mentioned at the end of Section \ref{subsubsec:humtrans_record_process}, after data cleaning and filtering, the remaining data volume is around 56.22 hours, which is the largest available humming dataset to date, shown in Table \ref{table:humtrans_comp}. Furthermore, apart from MIR-QBSH\cite{Jang}, we are one of the few humming datasets that provide music scores since MIDI files can be conveniently converted into sheet music using MuseScore 4. The duration distribution of our dataset is illustrated in Fig. \ref{fig:humtrans_dur_distribution}. The audio durations vary from 4.9s to 29.9s, with the majority falling within the range of 7s to 16s. The average duration is 13.9s.

\begin{table}[htbp]
\centering
\caption{Available humming dataset comparison.}
\begin{tabular}{l|c|c|c}
\hline\hline
Dataset                 & \#Hours & Score & Year       \\
\hline\hline
MTG-QBH \cite{salamon2013tonal}                 & 0.88    & no    & 2012       \\
AudioSet Humming \cite{gemmeke2017audio} & 1.20    & no    & 2017       \\
MIR-QBSH \cite{Jang}               & 9.85    & yes   & 2009  \\
MLEnd Hums and Whistles \cite{MLEnd} & 29.77   & no    & 2022       \\
\hline
\textbf{HumTrans (Ours)}         & \textbf{56.22}   & yes   & 2023       \\
\hline\hline
\end{tabular}
\label{table:humtrans_comp}
\end{table}

\section{Baseline Results}
\label{sec:baseline}

\noindent
In order to legally utilize the HumTrans dataset, we signed data authorization agreements with each student, and they were all aware that their collected humming data would be publicly released as part of the dataset. As mentioned earlier, the dataset consists of 14,614 audio files, which were partitioned into a training set (13,080 files), a validation set (765 files), and a test set (769 files) in a 90:5:5 ratio through random sampling. The evaluation metrics are precision, recall, and F1-score using the function \textit{transcription.precision\_recall\_f1\_overlap} from the \textit{mir\_eval} toolkit \cite{raffel2014mir_eval}. Following \cite{donahue2022melody}, a slight modification was performed on the evaluation script to achieve octave-invariant evaluation, which means a predicted transcription may receive full credit if it is off by a fixed octave shift but otherwise identical to the reference. This is closer to practical situations since individuals may have different vocal ranges. It is sufficient to ensure that the transcribed notes are correct within a fixed octave range, such as the C4 octave group. Modifying the overall key is a trivial task since it can be easily achieved using commonly available tuning software.

\begin{table}[htbp]
\label{table:humtrans_baseline_comp}
\centering
\caption{Baseline experimental results of four different vocal melody transcription models on our HumTrans dataset. The ``P'' stands for precision and ``R'' is recall.}
\scalebox{0.9}{\begin{tabular}{c|c|c|c|c|c|c}
\hline\hline
\multirow{2}*{Model} & \multicolumn{3}{c|}{Valid Set} & \multicolumn{3}{c}{Test Set} \\
\cline{2-7}
~ & P & R & F1 & P & R & F1 \\
\hline\hline
VOCANO \cite{vocano}    & 3.270    & 3.134 & 3.194 & 3.384    & 3.329 & 3.352 \\
Sheet Sage \cite{donahue2022melody} & 2.757    & 2.656 & 2.702 & 3.039    & 2.982 & 3.005 \\
MIR-ST500 \cite{wang2021preparation} & 6.258    & 6.448 & 6.341 & 5.686    & \textbf{5.853} & \textbf{5.755} \\
JDC-STP \cite{kum2022pseudo}   & \textbf{6.777}    & \textbf{6.785} & \textbf{6.741} & \textbf{5.844}    & 5.620 & 5.667 \\
\hline\hline
\end{tabular}}
\label{table:humtrans_baseline_res}
\end{table}

We present baseline results of four SOTA vocal melody transcription models on both validation and test sets of our HumTrans dataset, including VOCANO \cite{vocano}, Sheet Sage \cite{donahue2022melody}, MIR-ST500 \cite{wang2021preparation}, and JDC-STP \cite{kum2022pseudo}, shown in Table \ref{table:humtrans_baseline_res}. Among these models, MIR-ST500 is a singing dataset collected by \cite{wang2021preparation}, and they provided an AST baseline model using EfficientNet-b0 \cite{tan2019efficientnet}. We also included this model in our comparative experiments. For all the experiments, we directly utilized the pre-trained models provided by the authors to generate predicted transcription and compared them with the reference MIDI files. From Table \ref{table:humtrans_baseline_res}, we can observe that although JDC-STP performed slightly better than the other models, the transcription capabilities of all the models are still far from satisfactory. Therefore, there is significant room for improvement in the domain of humming melody transcription.

\section{Conclusion}
\label{sec:conclusion}

\noindent
In this paper, we introduced the HumTrans dataset which is primarily designed for humming melody transcription. To the best of our knowledge, this is the largest publicly available humming dataset to date consisting of solely hummed melodies. All recording files are accompanied with reference MIDI files using the same naming convention. With its extensive potential for downstream applications, such as music generation systems based on humming input, the HumTrans dataset opens up new possibilities for interactive music composition and creative exploration.

\section{Acknowledgements}
\label{sec:acknowledgements}

\noindent
I would like to extend my sincere appreciation to Xuanyi Ma, Yacheng Yang, Wenzhao Zhang and Zihao Fu for their invaluable assistance in connecting with students capable of participating in the humming project. I would also like to thank Jingjing Zhang and the other 9 students who greatly contributed to the completion of the humming dataset.


\small\bibliographystyle{IEEEbib}
\bibliography{strings}

\end{document}